\documentclass[10pt,fleqn,a4paper,twoside]{article}

\usepackage[sort&compress,numbers]{natbib}
\usepackage[margin=1in]{geometry}
\usepackage{amsmath,amssymb,amsfonts}
\usepackage{chemformula}
\usepackage{siunitx}
\usepackage{booktabs}
\usepackage{authblk}
\usepackage[english]{babel}
\usepackage[labelformat=simple]{subcaption}

\newcommand{\Reyn}{\operatorname{\mathit{Re}}}
\newcommand{\ReynCell}{\operatorname{\mathit{Re}}_{cell}}
\newcommand{\Mach}{\operatorname{\mathit{Ma}}}

\newcommand{\Knud}{\operatorname{\mathit{Kn}}}
\newcommand{\Lewi}{\operatorname{\mathit{Le}}}

\begin{document}

\title{
	INFLUENCE OF THE CONTROL TEMPERATURE OF PARK'S TWO-TEMPERATURE MODEL ON THE MARS PATHFINDER REACTIVE HYPERSONIC FLOW
}

\author[1]{{Gibson De Marchi Poltronieri} \thanks{Corresponding author: gibsondemarchi@gmail.com}}
\affil[1]{\small{Instituto Tecnológico de Aeronáutica -- DCTA\slash ITA, 12228-900 -- São José dos Campos, SP, Brazil}}

\author[1]{{Farney C. Moreira} \thanks{farney.coutinho@gmail.com}}

\author[2]{{João Luiz F. Azevedo} \thanks{joaoluiz.azevedo@gmail.com}}
\affil[2]{\small{Instituto de Aeronáutica e Espaço -- DCTA\slash IAE\slash ACE-L, 12228-904 -- São José dos Campos, SP, Brazil}}

\date{}

\maketitle

\begin{abstract}
	Numerical simulations of reactive hypersonic flow under thermodynamic and chemical non-e\-qui\-lib\-ri\-um conditions are presented for the Mars Pathfinder capsule. An 8-species chemical model is employed to simulate Mars' atmosphere. Park's two-temperature model is used to account for the thermal non-equilibrium phenomena. The present work analyzes the impact of different values of the weight factors used in Park's model, aiming to broaden the understanding of the weight factors influence. The code used to simulate the flows solves the Navier-Stokes equations modified to account for reacting gas mixtures. The findings are depicted in terms of the Mach number and temperature modes along the stagnation streamline in a region close to the shock wave. The present analysis also includes results regarding the stagnation point convective heat flux. The results indicate that varying the weight factors yields negligible differences in the shock wave position and stagnation point convective heat flux. The changes in the weight factors cause variations in the maximum temperature mode values in the non-equilibrium region. The results presented are in good agreement with experimental data present in the literature. The present work indicates that Park's two-temperature model weight factors can substantially affect the temperature mode distributions in the flow non-equilibrium region.

		\textit{CFD},
		\textit{Thermochemical non-equilibrium},
		\textit{Heat transfer},
		\textit{Chemical reactions}
\end{abstract}

\section{INTRODUCTION}\label{sec:introduction}

In recent years, public and private agencies have reignited the investments in space exploration. In a space mission, the entry/reentry phase is of utmost importance to guarantee the safety and integrity of the vehicle and its payload. The entry/reentry vehicle can achieve high speeds in the order of magnitude of \unit{\km\per\s}\@. The vehicle must withstand travel through varying density flow regimes, experiencing a high-temperature shock layer induced by the formation of a detached strong bow shock wave. The complexity of the shock wave and inside the shock layer increases considerably compared to the freestream flow. High-temperature gas phenomena become more evident in those conditions, such as thermodynamic and chemical non-equilibrium. Chemical reactions, such as dissociation, recombination, and ionization, occur, leading to changes in the gas mixture and impacting the gas mixture physicochemical properties. The project of hypersonic vehicles must account for those many complex phenomena accordingly to design a suitable thermal protection system (TPS).

The present work aims to evaluate the influence of Park's two-temperature model weight factors on the flow behavior of Mars Pathfinder hypersonic flow conditions. The hypersonic flows studied in the present work are simulated using the {LeMANS} parallel code \citep{lc_scalabrin_phd_2007}. The results obtained from the simulation are compared with existing numerical and experimental data present in the literature. The experimental data used as reference is the ``Run 749'' in \citet{hollis_1996}. The computations performed in this study consider an 8-species chemical model that simulates Mars' atmosphere. Moreover, this work assumes two gas mixture compositions, one purely of carbon dioxide, \ch{CO2}, and the other a mixture \SI{95}{\percent} of \ch{CO2} with \SI{5}{\percent} of molecular nitrogen, \ch{N2}\@.

The present work uses a set of weight factors that encompass the values regarded as good choices by the literature \citep{c_park_2010,niu_et_al_2018,moreira_wolf_azevedo_scitech_2021}. This study presents the simulated results in terms of the Mach number and temperature modes along the stagnation streamline, focusing on the non-equilibrium region near the shock wave. Moreover, the present analysis includes data for the stagnation point convective heat flux. Therefore, this work aims to broaden the knowledge about the influence of Park's two-temperature model weight factors on the Mars Pathfinder reactive hypersonic flow behavior.

\section{THEORETICAL FORMULATION}

\subsection{General Considerations}

The freestream density is a function of the vehicle altitude in the atmosphere. Therefore, the entry/reentry vehicle experiences different types of flow regimes throughout the entry/reentry trajectory. The Knudsen number is a dimensionless parameter often employed to differentiate flow regimes in continuum, transition, and free-molecule. This classification is of utmost importance because it influences which type of physic-mathematical formulation must be used to model the flow \citep{boyd_schwartzentruber_2017,agrawal_2020}. The Knudsen number is defined as
\begin{equation}
	\Knud = \lambda / L 
	\qquad \text{,}
	\label{eqn:knudsen_number_discrete}
\end{equation}
where \( \lambda \) is the mean free path, which means the average distance that a particle (atom or molecule) travels between successive collisions and \( L \) is the characteristic length, which is a representative measure of the dimension of the fluid-immersed body \citep{ga_bird_1994}. The Knudsen number can also be written as a function of the Reynolds number, Mach number, and ratio of specific heats as
\begin{equation}
	\Knud = \frac{\Mach}{\Reyn} \sqrt{\frac{\gamma \pi}{2}}
	\qquad \text{,}
	\label{eqn:knudsen_number_continuum}
\end{equation}
by assuming the hard-sphere (H-S) model for a particle collision cross section \citep{boyd_schwartzentruber_2017}. In the above equation, \( \gamma \) is the ratio of specific heats, \( \Mach \) is the Mach number, and \( \Reyn \) is the Reynolds number.

As stated, the Knudsen number is used to differentiate flow regimes. Therefore, sufficiently low Knudsen number values represent the continuum regime, while very high Knudsen number values imply the free-molecule regime. The literature often proposes varying Knudsen number values to differentiate these flow regimes. \citet{boyd_schwartzentruber_2017} state that the transition regime lies between \( \num{e-02} \leq \Knud < \num{e+01} \), whereas \citet{anderson_2006} defines the transition regime within \( \num{3e-02} \leq \Knud < 1 \). \citet{ga_bird_2013} defines the transition regime as the region between the slip flow and the free-molecule flow regimes. These differences indicate that the limiting Knudsen number values that define the transition flow regime remain unsettled in the literature.

The well-known Navier-Stokes equations can accurately model the continuum regime, including high-enthalpy flows. Additionally, the Navier-Stokes equations are also suited to model slip flow regimes. However, for Knudsen numbers within the transition and free-molecule regime, the Navier-Stokes equations are inadequate for capturing the flow physics accurately. In these cases, higher-order continuum transport formulations may be used to extend the viability of continuum formulations for higher Knudsen values, such as the Burnett, super-Burnett, Grad, and Onsager-Burnett equations \citep{agrawal_2020}. An alternative method for simulating flows for the transition and free-molecule regimes is through a discrete formulation. The Boltzmann equation and its variants are widely employed to describe the transition regime. The collisionless Boltzmann equation, for example, is appropriate for solving the free-molecule regime, where energy changes may occur without collisions between molecules. Moreover, the discrete approach is not limited to high-Knudsen numbers and can also accurately represent the continuum regime.

The hypersonic entry flows addressed in the present study have characteristic freestream Knudsen number of order of magnitude of \num{e-04}. Therefore, the flows studied are within the continuum regime, allowing the use of the Navier-Stokes equations. Furthermore, the solver employs Park's two-temperature model to account for thermodynamic non-equilibrium and weak ionization effects \citep{c_park_1989,lc_scalabrin_phd_2007}. Park's model couples the temperatures of the translational and rotational energy modes into a single translational-rotational temperature mode called \( T_{tr} \). The second temperature, namely \( T_{ve} \), describes the vibrational-electronic temperature mode associated with the coupling of the vibrational and electronic energy modes, along with the electron energy \citep{lc_scalabrin_phd_2007}.

\subsection{Conservation Equations and Related Models}

The present analysis employs the Navier-Stokes equations for multicomponent gas flows with source terms to account for chemical reactions, \( S_{c,v} \), and axisymmetric flow configuration, \( S_{axi} \). Moreover, the present formulation uses Park's two-temperature model to account for the non-equilibrium and energy transfer between the temperature modes considered. Therefore, the Navier-Stokes equations can be written in a multi-dimensional form using the index notation as 
\begin{equation}
	\frac{\partial Q}{\partial t}
	+ \frac{\partial (F_{j} - F_{v_{j}})}{\partial x_{j}}
	= S_{c,v} + S_{axi}
	\qquad \text{,}
	\label{eqn:navier_stokes}
\end{equation}
where \( Q \) is the vector of conservated variables
\begin{equation}
	Q = 
	\begin{Bmatrix}
		\rho_{1} & \dots & \rho_{ns} & \rho u_{i} & E & E_{ve}
	\end{Bmatrix}^{\mathsf{T}}
 \qquad \text{.}
	\label{eqn:conservated_variables}
\end{equation}
In the index notation, a free index represents a vectorial equation and repeated indices indicate a summation. Therefore, in the definition of \( Q \), \( \rho_{1},\ \dots,\ \rho_{ns} \) are the density of the \( ns \) chemical species, \( \rho \) is the gas mixture density, \( u_{i} \) are the velocity components, \( E \) is the total energy per unit volume of the gas mixture, and \( E_{ve} \) is the vibrational-electronic energy per unit volume of the gas mixture.

The inviscid and viscous flux terms, \( F_{j} \) and \( F_{v_{j}} \), respectively, are defined as 
\begin{equation}
	F_{j} = \begin{Bmatrix}
		\rho_{1} u_{j} \\
		\vdots \\
		\rho_{ns} u_{j} \\
		\rho u_{i} u_{j} + p \delta_{ij} \\
		\left(E + p\right) u_{j} \\
		E_{ve} u_{j}
	\end{Bmatrix}
	\qquad \qquad \text{and} \qquad \qquad
	F_{v_{j}} = \begin{Bmatrix}
		- J_{1,j} \\
		\vdots \\
		- J_{ns,j} \\
		\tau_{ij} \\
		\tau_{ij} - \left(q_{tr,j} + q_{ve,j}\right) - \sum \left(J_{s,j} h_{s}\right) \\
		- q_{ve,j} - \sum \left(J_{s,j} e_{ve,s}\right)
	\end{Bmatrix}
	\qquad \text{.}
\end{equation}
In the definition of \( F_{j} \), \( p \) is the gas mixture pressure and \( \delta_{ij} \) is the Kronecker delta function. In the definition of \( F_{v_{j}} \), \( J_{1,j},\ \dots,\ J_{ns,j} \) are the diffusion flux of the \( ns \) chemical species in the \( j \)-th direction, \( \tau_{ij }\) is the viscous stress tensor components, \( q_{tr,j} \) is the translational-rotational heat flux in the \( j \)-th direction, \( q_{ve,j} \) is the vibrational-electronic heat flux in the \( j \)-th direction, \( h_{s} \) is the enthalpy of the \( s \)-th chemical species, and \( e_{ve,s} \) is the specific vibrational-electronic energy of the \( s \)-th chemical species.

The diffusion flux, \( J_{ns,j} \), is given by Fick's Law as
\begin{equation}
	J_{s \neq e,j} = \rho D_{s} \frac{\partial Y_{s}}{\partial x_{j}}
	\qquad \text{,}
\end{equation}
where \( D_{s} \) is the diffusion coefficient and \( Y_{s} \) is the mass fraction of the \( s \)-th chemical species, except the electron. The diffusion flux for the electrons is calculated by the consideration of an ambipolar diffusion to guarantee charge neutrality \citep{lc_scalabrin_phd_2007}. Considering that Dalton's Law of partial pressure is valid and that each chemical species behaves as an ideal gas, the gas mixture pressure, \( p \), can be written as 
\begin{equation}
	p = 
	\sum\limits_{s \neq e} \left(\frac{\rho_{s} R_{u}}{M_{s}} T_{tr}\right)
	+ \frac{\rho_{e} R_{u}}{M_{e}} T_{ve}
	\qquad \text{,}
\end{equation}
where \( R_{u} \) is the universal gas constant \citep{gillespie_1930}. The viscous stress tensor, for a Newtonian fluid, is given by
\begin{equation}
	\tau_{ij} 
	= \mu \left(
		\frac{\partial u_{i}}{\partial x_{j}}
		+ \frac{\partial u_{j}}{\partial x_{i}}
	\right)
	- \left(\frac{2}{3}\mu - \beta\right) 
	\frac{\partial u_{k}}{\partial x_{k}} \delta_{ij}
	\qquad \text{,}
\end{equation}
where \( \mu \) is the mixture coefficient of viscosity and \( \beta \) is the bulk viscosity. The bulk viscosity arises from the momentum exchange between colliding molecules and their internal degrees of freedom, contributing to the dilatational term that appears in the normal stress. Therefore, the bulk viscosity could directly impact the calculations in the present formulation, especially in carbon dioxide flows. The literature presents some methods to calculate the bulk viscosity under different temperature ranges \citep{cramer_2012,jaeger_matar_muller_2018,sharma_kumar_2023}. However, those models are usually for low-temperature values. Therefore, the present formulation assumes the Stoke's hypothesis, \( \beta = 0 \). This assumption aims to avoid inaccuracies during the bulk viscosity calculations, which could consequently degrade the numerical results obtained. The present formulation employs Fourier's Law to model the heat fluxes as
\begin{equation}
	q_{tr,j} = - \kappa_{tr} \frac{\partial T_{tr}}{\partial x_{j}}
	\qquad \qquad \text{and} \qquad \qquad
	q_{ve,j} = - \kappa_{ve} \frac{\partial T_{ve}}{\partial x_{j}}
	\qquad \text{,}
\end{equation}
where \( \kappa_{tr} \) and \( \kappa_{ve} \) are the thermal conductivity coefficients of the gas mixture associated with the translational-rotational and vibrational-electronic temperature modes, respectively.

This work employs two approaches to calculate the transport properties \citep{lc_scalabrin_phd_2007}. The first approach uses \citet{wilke_1950} semi-empirical mixing rule, \citet{blottner_johnson_ellis_1971} curve fits, and Eucken's relation \citep{vincenti_kruger_1982,anderson_2006}. The second approach is the one proposed by \citet{gupta_yos_thompson_lee_1990}. The first approach is suited for low velocity flows with maximum temperatures around \SI{10000}{\kelvin} and is not designed for ionized flows. The second model is suitable for high-speed, around \SI{10}{\km\per\s}, and weakly ionized flows. Considering the first approach, \citet{wilke_1950} semi-empirical mixing rules used to calculate the gas mixture viscosity coefficient and the thermal conductivity coefficient are given by
\begin{equation}
	\mu = \sum\limits_{s}^{ns} \frac{X_{s} \mu_{s}}{\phi_{s}}
 	\qquad \qquad \text{and} \qquad \qquad
	\kappa = \sum\limits_{s}^{ns} \frac{X_{s} \kappa_{s}}{\phi_{s}}
 	\qquad \text{,}
\end{equation}
respectively. In the above equation, the \( \phi_{s} \) term is given by
\begin{equation}
	\phi_{s} = 
	\sum\limits_{r}^{nr} X_{r}
 	\left[
		1 + \sqrt{\frac{\mu_{s}}{\mu_{r}}}
 		\left(\frac{M_{r}}{M_{s}}\right)^{1/4}
 	\right]^{2}
 	\left[
 		\sqrt{8 \left(1 + \frac{M_{s}}{M_{r}}\right)}
 	\right]^{-1}
 	\qquad \text{,}
\end{equation}
where \( X \) is the molar fraction and \( M \) is the molar weight of the \( s \)-th and \( r \)-th chemical species. The chemicals species dynamic viscosity coefficient, \( \mu_{s} \), is calculated by Blottner's curve fits as
\begin{equation}
	\mu_{s} = 0.1 \exp\left\{
 		\left[A_{s} \ln(T) + B_{s}\right]
 		\ln(T) + C_{s}
 	\right\}
 	\qquad \text{,}
\end{equation}
where \( A_{s} \), \( B_{s} \), and \( C_{s} \) are constants for each chemical species \citep{blottner_johnson_ellis_1971}. The thermal conductivity coefficients, \( \kappa_{tr} \) and \( \kappa_{ve} \), are calculated for each chemical species by Eucken's relation as
\begin{equation}
	\kappa_{tr,s} = \mu_{s} \left( \frac{5}{2} C_{v_{t,s}} + C_{v_{r,s}} \right) 
 	\qquad \qquad \text{and} \qquad \qquad
	\kappa_{ve,s} = \mu_{s} C_{v_{ve,s}}
 	\qquad \text{,}
\end{equation}
where \( C_{v} \) is the specific heat at constant volume for different internal energy modes of different chemical species \citep{vincenti_kruger_1982,anderson_2006}. The diffusion coefficient for each chemical species, \( D_{s} \), is calculated simply by a single binary coefficient \( D \), which ensures that the sum of the diffusion fluxes is zero. Thus, given by
\begin{equation}
 	D = \frac{\kappa_{tr}}{\rho C_{p_{tr}}} \Lewi
 	\qquad \text{,}
\end{equation}
where \( C_{p_{tr}} \) is the gas mixture specific heat at constant pressure associated with the translational-rotational temperature mode and \( \Lewi \) is the Lewis number, a dimensionless number that represents the ratio between thermal diffusivity and mass diffusivity. However, this approach is not accurate for velocities above \SI{10}{\km\per\s} \citep{lc_scalabrin_phd_2007}.

In Eq.~(\ref{eqn:navier_stokes}), the source term, \( S_{axi} \), represents the additional surface stress that appears by an axisymmetric formulation. The contribution is only to the y-momentum equation to counterbalance the pressure and viscous forces acting on the side of the surface of the control volume. Therefore, the \( S_{axi} \) is given by
\begin{equation}
	S_{axi} = 
	\begin{Bmatrix}
		0 & \dots & 0 & 0 & \left[
			-p + 2\mu \left(
				\dfrac{u_{2}}{\bar{x}_{2}}
				- \dfrac{1}{3} \dfrac{\partial u_{k}}{\partial x_{k}}
			\right)
		\right] \dfrac{\delta_{i2}}{\bar{x}_{2}} & 0 & 0
	\end{Bmatrix}^{\mathsf{T}}
	\qquad \text{,}
\end{equation}
where 
\begin{equation}
	\frac{\partial u_{k}}{\partial x_{k}}
	= \frac{\partial u_{1}}{\partial x_{1}}
	+ \frac{\partial u_{2}}{\partial x_{2}}
	+ \frac{u_{2}}{\bar{x}_{2}}
	\qquad \text{.}
\end{equation}
In the above formulation, the \( \bar{x}_{2} \) is the radial coordinate measure from the axis of symmetry to the cell centroid and \( x_{1} \) and \( x_{2} \) are the axial and radial directions, respectively.

The source term \( S_{c,v} \) is associated with the rate of production or consumption of mass of the chemical species in the reacting flow and the energy related to the vibrational energy mode. Therefore, the \( S_{c,v} \) source term is written as
\begin{equation}
	S_{c,v} = 
	\begin{Bmatrix}
		\dot{w}_{1} & \dots & \dot{w}_{ns} & 0 & 0 & 0 & 0 & \dot{w}_{v}
	\end{Bmatrix}^{\mathsf{T}}
	\qquad \text{,}
\end{equation}
where \( \dot{w}_{1},\ \dots,\ \dot{w}_{ns} \) are the rate of mass production of the \( ns \) chemical species and \( \dot{w}_{v} \) is the vibrational energy source term.

\subsection{Chemical Species and Model of Chemical Reactions}

High-enthalpy flows, such as those encountered in hypersonic atmospheric entry/reentry flows, cause changes in the gas mixture composition due to chemical reactions. These changes also impact the physicochemical properties of the flow. Therefore, chemical models have been developed to represent those high-enthalpy flow phenomena according to the flow physicochemical complexity \citep{gnoffo_gupta_shinn_1989}. The present work uses an 8-species chemical model to simulate Mars' atmosphere. This chemical model species consists of \ch{CO2}, \ch{CO}, \ch{N2}, \ch{O2}, \ch{NO}, \ch{N}, \ch{O}, and \ch{C}.

The present work considers a finite-rate chemistry model for the reacting gas mixture of the reactive hypersonic flows studied \citep{lc_scalabrin_phd_2007}. The chemical reactions are generically given by
\begin{equation}
	\sum \alpha_{s,r} \left[S\right] \ch{<=>} \sum \beta_{s,r} \left[S\right]
	\qquad \text{,}
\end{equation}
where \( \left[S\right] \) represents the chemical species listed in the model and \( \alpha_{s,r} \) and \( \beta_{s,r} \) are the stoichiometric coefficients of the \( s \)-th chemical species that balance the \( r \)-th chemical reaction of the model. The present work standardizes the forward chemical reactions as endothermic reactions and the backward chemical reactions as exothermic reactions. Endothermic reactions absorb energy from the surroundings while exothermic reactions release energy to the surroundings \citep{atkins_paula_2006}.

The rate of production or consumption of the \( s \)-th chemical species is given by
\begin{equation}
	\dot{w}_{s} = 
	M_{s} \sum\limits_{r}^{nr} 
	\left(\beta_{s,r} - \alpha_{s,r}\right) 
	\left[
		k_{f,r} \prod\limits_{s}^{ns} 
		\left(
			\frac{\rho_{s}}{M_{s}}
		\right)^{\alpha_{s,r}}
		-
		k_{b,r} \prod\limits_{s}^{ns} 
		\left(
			\frac{\rho_{s}}{M_{s}}
		\right)^{\beta_{s,r}}
	\right]
	\qquad \text{,}
\end{equation}
where \( nr \) represents the number of chemical reactions that the \( s \)-th chemical species participates, and \( k_{f,r} \) and \( k_{b,r} \) are the forward and backward chemical reaction rates of the \( r \)-th chemical reaction, respectively.

The thermodynamic and chemical non-equilibrium phenomena occur when the characteristic time scale of the flow is in the same order of magnitude as the relaxation time of the physicochemical phenomenon processes. The thermodynamic and chemical non-equilibrium in the flow affects the forward and backward chemical reaction rates. The present work uses \citet{c_park_1989} two-temperature model to account for the thermal non-equilibrium. In Park's model, the translational and rotational temperature modes are coupled into the translational-rotational temperature mode, \( T_{tr} \), and the vibrational, electronic, and electron translational temperature modes are coupled into the vibrational-electronic temperature mode, \( T_{ve} \). Moreover, Park's model defines the control temperature, \( T_{c} \), as a combination of \( T_{tr} \) and \( T_{ve} \), given by \( T_{c} = T_{tr}^{a} T_{ve}^{b} \), where \( a \) and \( b = 1 - a \) are weight factors that control the energy transfer between dissociation and ionization reactions. It must be noted that the control temperature, \( T_{c} \), can assume the value of \( T_{tr} \), \( T_{ve} \), or \( T_{tr}^{a} T_{ve}^{b} \) depending on the of chemical reaction \citep{lc_scalabrin_phd_2007,niu_et_al_2018}. This formulation accounts for the fact that vibrationally excited molecules are more likely to dissociate \citep{c_park_1989}. The typical values for \( a \) and \( b \) indicated by the literature are \( a = b = 0.5 \) or \( a = 0.7,\ b = 0.3 \) \citep{lc_scalabrin_phd_2007,c_park_2010}.

The forward reaction rate, \( k_{f} \), is a function of the control temperature, \( T_{c} \), and is calculated by Arrhenius curve fits of the form
\begin{equation}
	k_{f,r} (T_{c})
	= C_{f,r} T_{c}^{\eta_{r}} \exp\left(- \theta_{r} / T_{c}\right)
	\qquad \text{,}
\end{equation}
where \( C_{f,r} \), \( \eta_{r} \), and \( \theta_{r} \) are constants from \citet{c_park_1990}. The backward reaction rate is given by
\begin{equation}
	k_{b,r} (T_{c}) = k_{f,r} (T_{c}) / k_{eq,r} (T_{c})
	\qquad \text{,}
\end{equation}
where \( k_{eq,r} \) is the equilibrium constant of the \( r \)-th chemical reaction \citep{atkins_paula_2006}. The equilibrium constant, \( k_{eq} \), is calculated by curve fits as follows
\begin{equation}
	k_{eq,r} (T_{c})
	=
	\exp \left[
		A_{1} \left(\frac{T_{c}}{10^{4}}\right)
		+ A_{2}
		+ A_{3} \ln \left(\frac{10^{4}}{T_{c}}\right)
		+ A_{4} \left(\frac{10^{4}}{T_{c}}\right)
		+ A_{5} \left(\frac{10^{4}}{T_{c}}\right)^{2}
	\right]
	\qquad \text{,}
\end{equation}
where the coefficients \( A_{1} \), \( A_{2} \), \( A_{3} \), \( A_{4} \), and \( A_{5} \) are functions of the local number density of the flow within the range of data tabulated in \citet{c_park_1989}. In cases where the number density value is outside the tabulated data, the formulation uses the maximum and minimum values of the tabulated data accordingly. This approach may create numerical instabilities and errors, particularly in edge cases.

\section{NUMERICAL FORMULATION}

The present work uses the ``Le'' Michigan Aerothermodynamic Navier-Stokes Solver ({LeMANS}), a parallel code for unstructured meshes developed at the University of Michigan by \citet{lc_scalabrin_phd_2007}. The code solves the Navier-Stokes equations including chemical reactions and energy transfer between different energy modes with a finite volume method with a cell-centered approach. Furthermore, {LeMANS} is capable of handling axisymmetric flow configurations using meshes composed solely of quadrilaterals to better resolve the boundary layers and shock waves \citep{lc_scalabrin_phd_2007}.

{LEMANS} uses a modified Steger-Warming flux vector splitting (FVS) scheme to discretize the inviscid fluxes across the cell faces \citep{maccormack_candler_1989}. The modified method switches to the original Steger-Warming FVS scheme in the vicinity of shock waves or other discontinuities by the action of a pressure switch \citep{steger_warming_1981}. According to \citet{maccormack_candler_1989}, the original Steger-Warming FVS scheme is highly dissipative and suitable for high-gradient regions, such as shock waves. The modified Steger-Warming FVS scheme is less dissipative and suitable to solve boundary layer-type flows. Moreover, {LeMANS} implements a second-order reconstruction scheme for the inviscid fluxes \citep{lc_scalabrin_phd_2007}.

The present formulation employs a second-order centered scheme to calculate the viscous fluxes at the cell faces using the property values of the centroid and nodes that compose the face. The values of the properties at the nodes are calculated using a simple average of the cell values of the cells that share the node. This approach increases the stencil employed in the derivative calculations, avoiding loss of accuracy in unstructured meshes. The current formulation employs a no-slip and non-catalytic isothermal wall boundary conditions for the wall-type surfaces. \citet{fc_moreira_phd_2020} presents a detailed analysis of the influence of catalytic and non-catalytic wall boundary conditions for the configuration used in this work. The axisymmetric source term is spatially discretized with the same approach used for the viscous terms. The present formulation calculates the values of properties on the left and right sides of a cell face using the value of the respective cell centroid and the nodes that compose the control volume \citep{jawahar_kamath_2000}.

Numerical instabilities may arise related to chemical reactions included by the chemical source term. The first problem that may appear is that the chemical reaction rates may achieve large values depending on the control temperature, \( T_{c} \), especially for the low equilibrium constant values \citep{c_park_1988}. Another problem that may arise is that the density of the chemical species can assume negative values during the convergence process of the solver. Negative density values yield negative values for the source terms, thus numerical instabilities. \citet{lc_scalabrin_phd_2007} proposed a modified temperature to overcome the problem related to the source term calculations.

Numerical instabilities may appear due to the use of explicit methods for the time integration of the Navier-Stokes equations with chemical source terms. Moreover, the time step restriction due to the stiffness of the formulation presented does not allow an adequate convergence rate to the solution \citep{hirsch_2007}. Implicit time integration schemes are suited for a stiff system of equations, allowing larger time steps while avoiding the growth of numerical instabilities. Therefore, the current formulation uses point and line implicit time integration schemes \citep{lc_scalabrin_phd_2007,venkatakrishnan_1995}.

\section{RESULTS AND DISCUSSION}

\subsection{Flow Conditions Considered in the Simulations}

This section presents the results for the entry flow over the Mars Pathfinder capsule regarding the impact of the weight factor values. This work considers two types of flow gas mixtures that simulate Mars' atmosphere. One of the gas mixtures is purely composed of carbon dioxide, \ch{CO2}, and the other is composed of \SI{95}{\percent}~(w/w) of \ch{CO2} and \SI{5}{\percent}~(w/w) of \ch{N2}. The (w/w) represents a mass fraction value notation. The freestream conditions and the flow composition are based on experimental data available for the Mars Pathfinder capsule, obtained in the HYPULSE expansion tube. Table~\ref{tab:mars_pathfinder_freestream_conditions} shows the freestream condition for each gas mixture considered in the present work. The name ``Run 749'' refers to a particular experiment performed by \citet{hollis_1996}.

\begin{table}[!htbp]
	\centering
	\caption{Mars Pathfinder ``Run 749'' freestream conditions}
	\label{tab:mars_pathfinder_freestream_conditions}
	\begin{tabular}{*{9}{c}}
		\toprule
		Gas Mixture &
		\( \rho_{\infty} \)  &
		\( T_{\infty} \) &
		\( T_{w} \) &
		\( U_{\infty} \) &
		\( R \) &
		\( \Mach_{\infty} \) &
		\( \Reyn_{\infty} \) & 
		\( \Knud_{\infty} \) \\
		&
		\( \left[\unit{\kg\per\cubic\m}\right] \)  &
		\( \left[\unit{\kelvin}\right] \) &
		\( \left[\unit{\kelvin}\right] \) &
		\( \allowbreak \left[\unit{\m\per\s}\right] \) &
		\( \left[\unit{\m}\right] \) &
		&
		& 
		\\
		\midrule
		\ch{CO2} 		& \num{5.75e-03} & \num{1045} & \num{300} & \num{4769}
			& \num{0.0254} & \num{9.89} & \num{2.03e+04} & \num{6.63e-04} \\
		\ch{CO2 + N2} 	& \num{5.75e-03} & \num{1045} & \num{300} & \num{4769}
			& \num{0.0254} & \num{9.89} & \num{2.03e+04} & \num{6.64e-04} \\
		\bottomrule
	\end{tabular}
\end{table}

In Tab.~\ref{tab:mars_pathfinder_freestream_conditions}, \( \rho_{\infty} \) is the freestream density, \( T_{\infty} \) and \( T_{w} \) are the freestream and wall temperatures, respectively, \( U_{\infty} \) is the freestream flow speed, \( R \) is the circumference radius of the thermal shield and is the reference length used to calculate the dimensionless coordinates \( X/R \) and \( Y/R \) used in the upcoming figures, \( \Mach_{\infty} \) is the freestream Mach number, \( \Reyn_{\infty} \) is the freestream Reynolds number, and \( \Knud_{\infty} \) is the freestream Knudsen number.

\subsection{Computational Grid}

The computational grid used for the numerical simulations performed in the present work is composed solely of quadrilaterals in axisymmetric configuration. The computational grid covers the thermal shield of the thermal protection system of the Mars Pathfinder capsule. In this computational grid, there are two core refinement regions. One of the regions is at the shock wave position, defined by two computational lines. This first refinement region aims to better capture the shock wave and property gradients due to non-equilibrium phenomena. Therefore, the present work calls this refinement region as the ``non-equilibrium region.'' The definition of the non-equilibrium region is based on previous numerical data from \citet{moreira_wolf_azevedo_scitech_2021,fc_moreira_phd_2020}. The second refinement region is near the vehicle wall. The mesh refinement near the wall allows better capturing of temperature gradients, aiming for the correct calculation of wall convective heat flux.

\begin{figure}[!htbp]
	\centering
	\begin{subfigure}{0.3\textwidth}
		\centering
		\includegraphics[height=5.5cm]{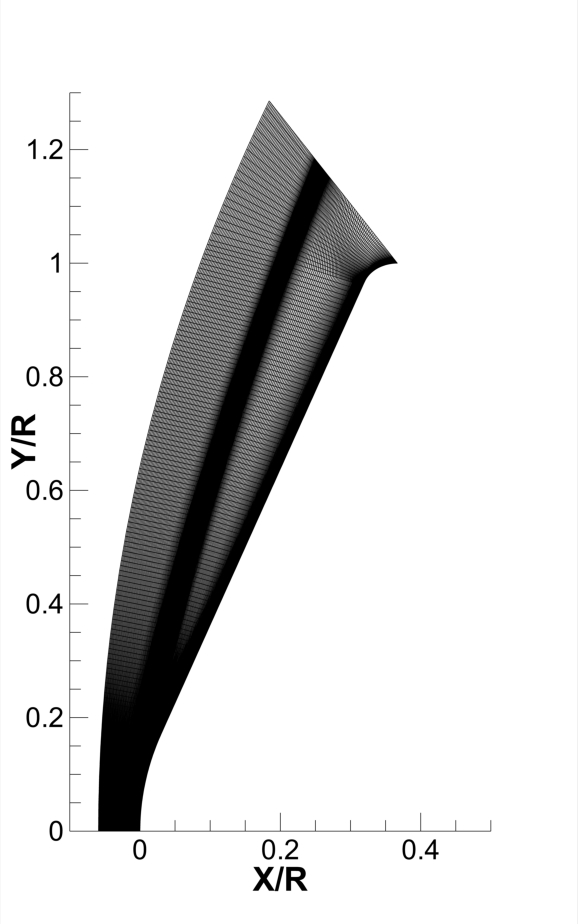}
		\caption{Computational domain.}
		\label{subfig:mars_full_mesh}
	\end{subfigure}
	\hspace{0.05\textwidth}
	\begin{subfigure}{0.48\textwidth}
		\centering
		\includegraphics[height=5.5cm]{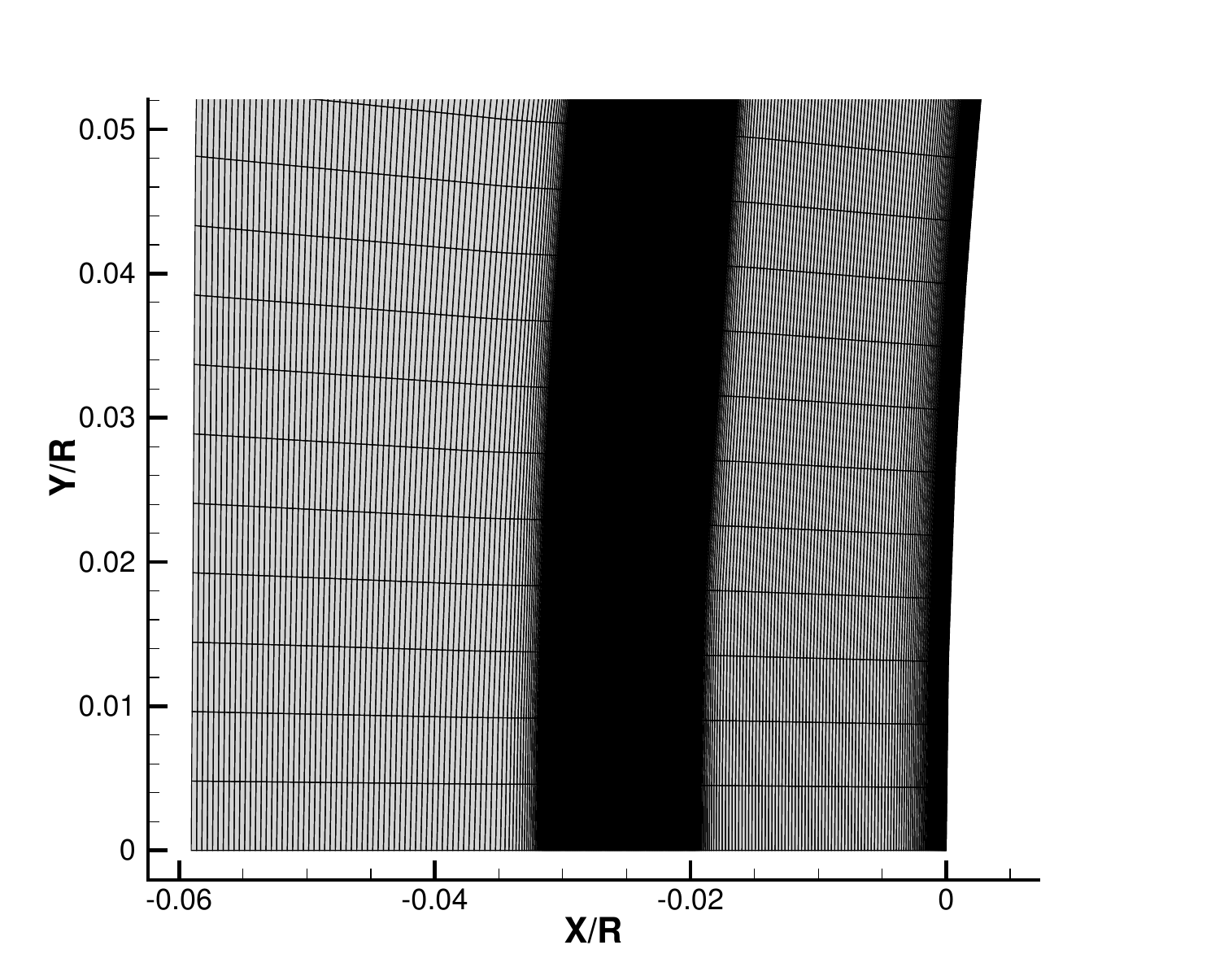}
		\caption{Computational domain near the stagnation line.}
		\label{subfig:mars_stag_mesh}
	\end{subfigure}
	\caption{Mars Pathfinder computational domains.}
	\label{fig:mars_pathfinder_meshes}
\end{figure}

Based on the work of \citet{moreira_wolf_azevedo_scitech_2023} and \citet{gm_poltronieri_msc_2024}, this study refines the non-equilibrium region uniformly using the cell Reynolds number, \( \ReynCell \), of approximately \num{1} and employs a stretching factor of \SI{10}{\percent} for the mesh transition to other mesh regions. The present work refines the mesh near the wall using a \( \ReynCell \approx 1 \) with a stretching factor of \SI{5}{\percent}. The smallest grid distance in the wall-normal direction at the vehicle wall is \( \Delta n = \SI{5e-07}{\m} \). The computational grid generated for the present work has \num{460} cells in the wall-normal direction and \num{160} cells in the streamwise direction. Figure~\ref{fig:mars_pathfinder_meshes} presents an example of the computational grid described.


\subsection{Shock Wave Position}

\begin{figure}[!htbp]
	\centering
	\begin{subfigure}{0.45\textwidth}
		\centering
		\includegraphics[height=5cm]{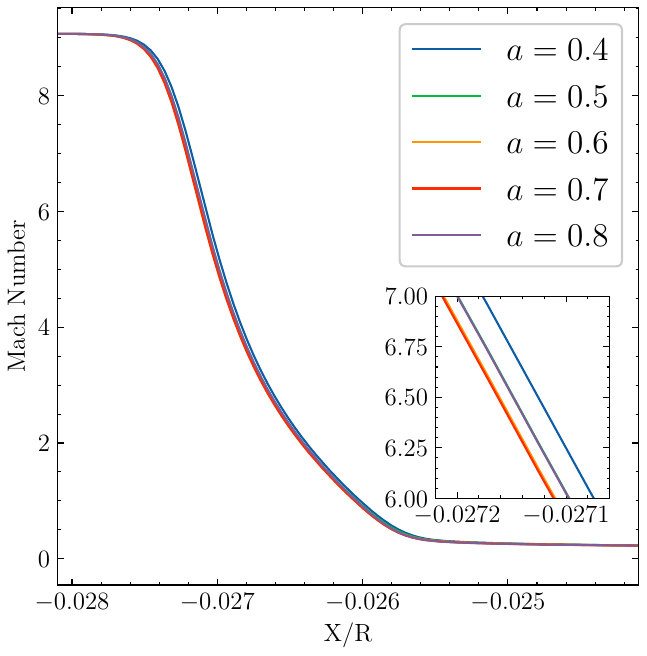}
		\caption{\ch{CO2} flow.}
		\label{subfig:mars_cmp_mach_co2}
	\end{subfigure}
	\hspace{0.05\textwidth}
	\begin{subfigure}{0.45\textwidth}
		\centering
		\includegraphics[height=5cm]{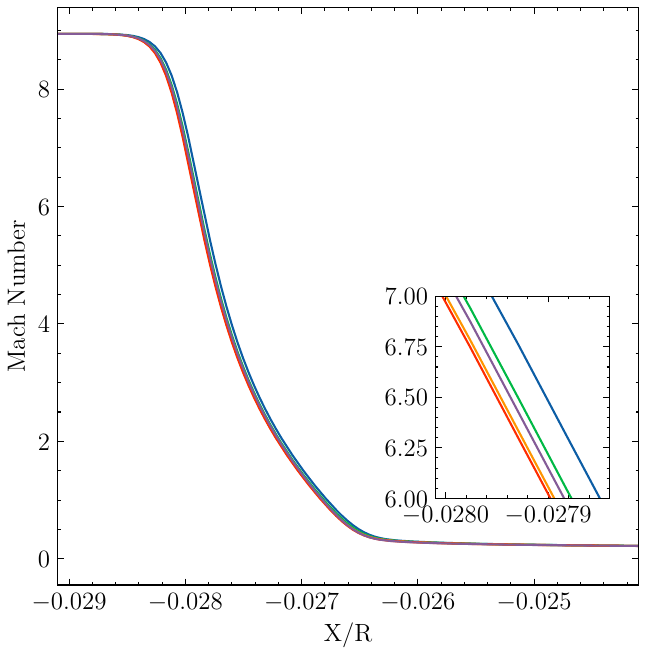}
		\caption{\ch{CO2 + N2} flow.}
		\label{subfig:mars_cmp_mach_co2_n2}
	\end{subfigure}
	\caption{Mach number distribution along the stagnation streamline indicating the shock wave position.}
	\label{fig:mars_cmp_shock_wave}
\end{figure}

Figure~\ref{fig:mars_cmp_shock_wave} shows the Mach number distribution along the stagnation streamline in the non-equilibrium region for each set of weight factors. Figure~\ref{subfig:mars_cmp_mach_co2} represents the results for the gas mixture composition of \ch{CO2} while Fig.~\ref{subfig:mars_cmp_mach_co2_n2} represents results for the gas mixture composition of \ch{CO2 + N2}. Note that the legend only presents the \( a \) weight factor because the other weight factor is \( b = 1 - a \)\@. The results for the shock wave position obtained in the present analysis are in good agreement with the numerical data from \citet{moreira_wolf_azevedo_scitech_2021} and \citet{fc_moreira_phd_2020}. The position of the shock wave seems to move slightly away from the vehicle wall, represented by \( X / R = 0 \), as the value of \( a \) increases. However, this behaviors breaks down for \( a = 0.7 \) and \( a = 0.8 \)\@. The differences observed in the shock wave position are significantly small compared to the stagnation line length. Therefore, the impact of the weight factors on the shock wave position can be considered as very small indeed.

The behavior observed in Fig.~\ref{fig:mars_cmp_shock_wave} follows the same behavior reported in \citet{poltronieri_moreira_azevedo_cobem_2023}, for a different geometry and quite different chemical compositions of the atmosphere, where the increase of the \( a \) weight factor value also yields shock waves positioned slightly away from the vehicle body. However, as presented in this section, the pattern breaks down for \( a = 0.7 \) and \( a = 0.8 \). Further analyses are still being performed in order to fully understand this reversion in the trend of the positioning of the shock wave with the increase of the \( a \) coefficient.


\subsection{Temperature Modes}

\begin{figure}[!htbp]
	\centering
	\begin{subfigure}{0.45\textwidth}
		\centering
		\includegraphics[width=0.9\textwidth]{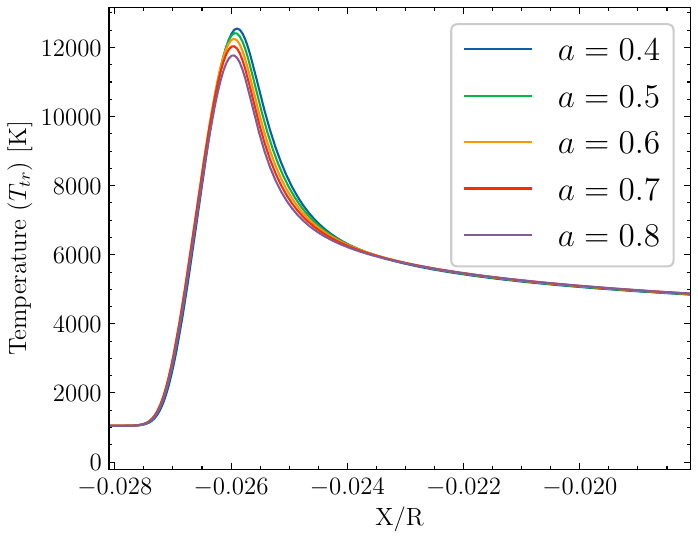}
		\caption{\( T_{tr} \) distribution for \ch{CO2} flow.}
		\label{subfig:mars_cmp_t_tr_co2}
	\end{subfigure}
	\hspace{0.05\textwidth}
	\begin{subfigure}{0.45\textwidth}
		\centering
		\includegraphics[width=0.9\textwidth]{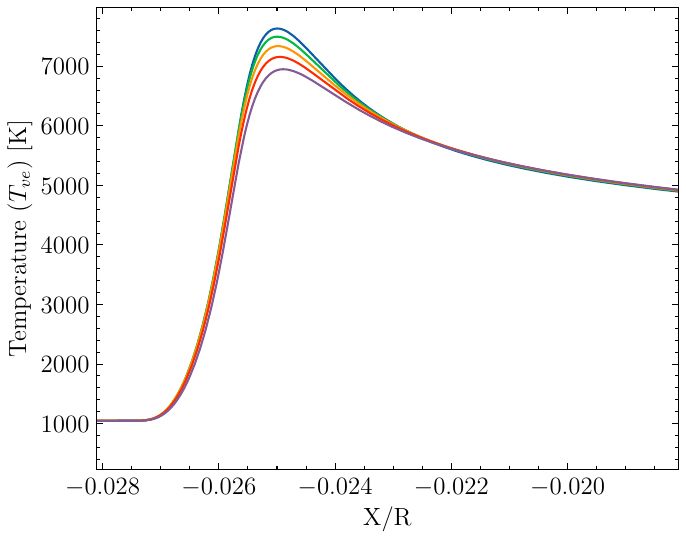}
		\caption{\( T_{ve} \) distribution for \ch{CO2} flow.}
		\label{subfig:mars_cmp_t_ve_co2}
	\end{subfigure}
	\\
	\begin{subfigure}{0.45\textwidth}
		\centering
		\includegraphics[width=0.9\textwidth]{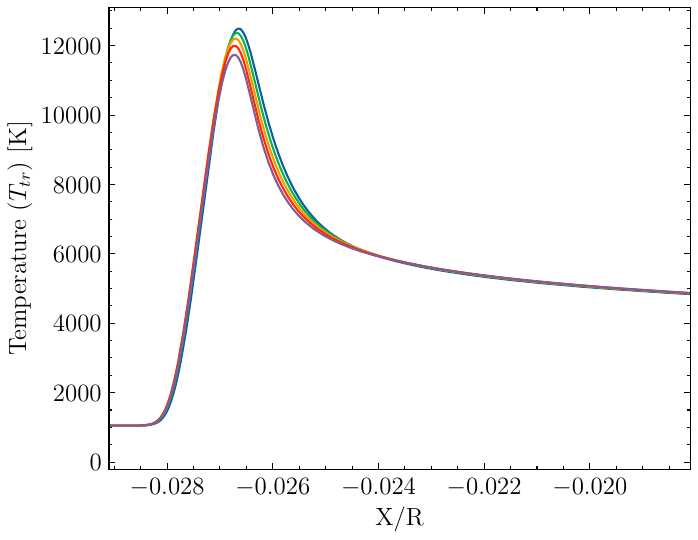}
		\caption{\( T_{tr} \) distribution for \ch{CO2 + N2} flow.}
		\label{subfig:mars_cmp_t_tr_co2_n2}
	\end{subfigure}
	\hspace{0.05\textwidth}
	\begin{subfigure}{0.45\textwidth}
		\centering
		\includegraphics[width=0.9\textwidth]{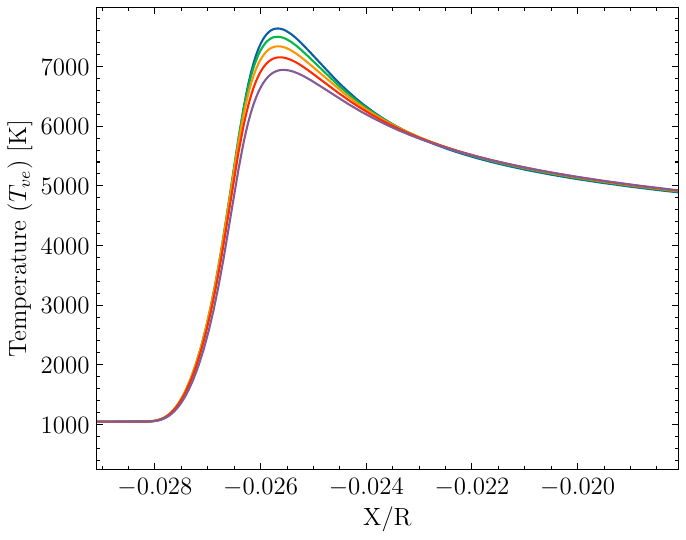}
		\caption{\( T_{ve} \) distribution for \ch{CO2 + N2} flow.}
		\label{subfig:mars_cmp_t_ve_co2_n2}
	\end{subfigure}
	\caption{Temperature mode distributions along the stagnation streamline.}
	\label{fig:mars_cmp_temperature_modes}
\end{figure}

Figure~\ref{fig:mars_cmp_temperature_modes} shows the \( T_{tr} \) and \( T_{ve} \) temperature mode distributions along the stagnation streamline in the non-equilibrium region. Figures~\ref{subfig:mars_cmp_t_tr_co2} and \ref{subfig:mars_cmp_t_ve_co2} are the temperature mode distributions for the \ch{CO2} flow and Figs.~\ref{subfig:mars_cmp_t_tr_co2_n2} and \ref{subfig:mars_cmp_t_ve_co2_n2} are the temperature mode distributions for the \ch{CO2 + N2} flow. Figures~\ref{subfig:mars_cmp_t_tr_co2} and \ref{subfig:mars_cmp_t_tr_co2_n2} represent the \( T_{tr} \) temperature mode and Figs.~\ref{subfig:mars_cmp_t_ve_co2} and \ref{subfig:mars_cmp_t_ve_co2_n2} represent the \( T_{ve} \) temperature mode.

The results for the \( T_{tr} \) and \( T_{ve} \) temperature modes show maximum temperature values larger than the values presented by \citet{fc_moreira_phd_2020}, \citet{moreira_wolf_azevedo_scitech_2021}, and \citet{lc_scalabrin_phd_2007}. This result is related to the mesh refinement in the non-equilibrium region as shown by \citet{gm_poltronieri_msc_2024}. In Fig.\ref{fig:mars_cmp_temperature_modes}, it is clearly seen that the changes in the weight factor values cause changes in both temperature modes. The increase of the \( a \) weight factor yields lower \( T_{tr} \) and \( T_{ve} \) temperature mode distributions. This behavior is expected because the \( a \) weight factor is related to the \( T_{tr} \) temperature mode, which has higher values than the \( T_{ve} \) temperature mode. Therefore, the control temperature, \( T_{c} \), becomes closer to the \( T_{tr} \) temperature distribution as the value of \( a \) increases. Thus, an increase in the \( T_{c} \) value yields higher forward reaction rates, \( k_{f} \), values for the dissociation reactions. Moreover, as the forward reactions are endothermic, they absorb energy from the surroundings. Therefore, the lower temperature distributions agree with the expected physical behavior. The maximum difference between the peak values for the \( T_{tr} \) temperature mode is around \SI{1000}{\kelvin}. For the \( T_{ve} \) temperature mode, the maximum difference between the peak values is around \SI{600}{\kelvin}.

\citet{niu_et_al_2018} present results of hypersonic flows over a blunt body configuration, named {BSUV-II}, where the authors compare three two-temperature models, including Park's two-temperature model. \citet{niu_et_al_2018} also tested two sets of weight factors, \( a = b = 0.5 \) and \( a = 0.7 \) and \( b = 0.3 \)\@. In their study, the authors observed that the change in the weight factors only impacted the vibrational-electronic temperature mode. However, the present study shows that the weight factors also impact the translational-rotational temperature mode. Note that the freestream conditions and vehicle body are different. Therefore, the flow conditions and body geometry for the Mars Pathfinder capsule may be sufficient to show the impact of the weight factors in both temperature mode distributions. \citet{poltronieri_moreira_azevedo_cobem_2023} also show that the weight factors of Park's two-temperature model impact the \( T_{tr} \) temperature mode for the {FIRE II} reentry capsule hypersonic flow conditions. The freestream, between the far-field boundary and the shock wave, and the shock layer regions show a thermodynamic equilibrium behavior. In these regions, the temperature modes assume the same values, \( T_{c} = T_{tr} = T_{ve} \), which is the expected behavior for the flow in equilibrium conditions.

\subsection{Convective Heat Flux}

\begin{figure}[!htbp]
	\centering
	\begin{subfigure}{0.45\textwidth}
		\centering
		\includegraphics[height=7cm]{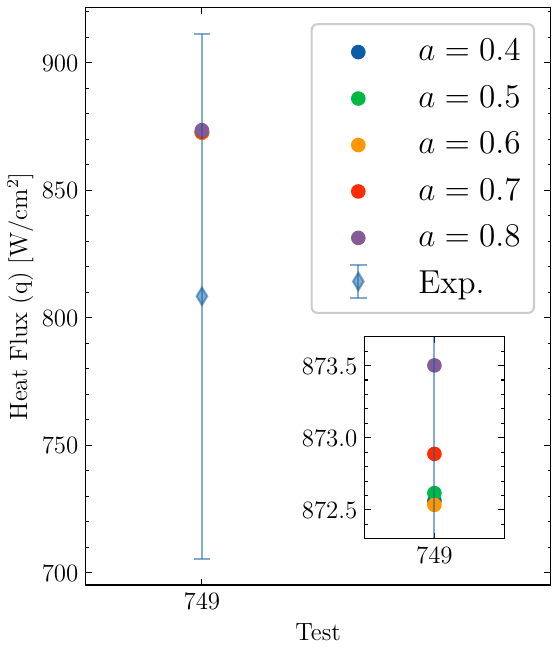}
		\caption{\ch{CO2} flow.}
		\label{subfig:mars_cmp_stag_q_co2}
	\end{subfigure}
	\hspace{0.05\textwidth}
	\begin{subfigure}{0.45\textwidth}
		\centering
		\includegraphics[height=7cm]{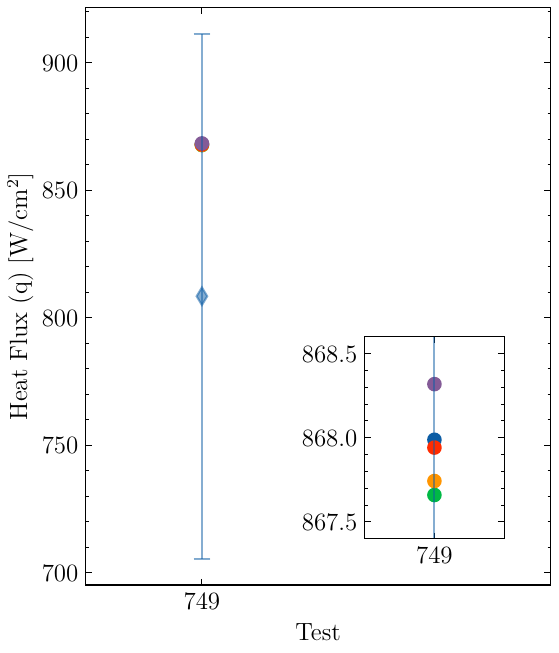}
		\caption{\ch{CO2 + N2} flow.}
		\label{subfig:mars_cmp_stag_q_co2_n2}
	\end{subfigure}
	\caption{Stagnation point convective heat flux values.}
	\label{fig:mars_cmp_stag_q}
\end{figure}

Figure~\ref{fig:mars_cmp_stag_q} shows the stagnation point convective heat flux for the sets of weight factors with \( a = 0.4 \), \( 0.5 \), \( 0.6 \), \( 0.7 \), and \( 0.8 \)\@. Figure~\ref{subfig:mars_cmp_stag_q_co2} presents the stagnation point convective heat flux results for the flow composed solely of \ch{CO2} and Fig.~\ref{subfig:mars_cmp_stag_q_co2_n2} presents the corresponding results for the flow of \ch{CO2 + N2}. The experimental data are from \citet{hollis_1996}. The margin of error of the experiments is reported as approximately \SI{12.7}{\percent} for the total stagnation point heat flux. The ``total'' stagnation point heat flux refers to the convective plus radiative stagnation point heat fluxes. The blue diamond-shaped symbol represents the experimental measurement value, while the blue bars represent the uncertainty of the experimental measurement.

The results obtained for each set of weight factors in both flow conditions show that the differences between the results are negligible. Note that both figures have an inset that presents a zoom of the present simulation results to indicate how close these calculated values of convective wall heat transfer are. The differences between the results for each set of weight factors are less than \SI{0.15}{\percent} of the averaged predicted convective heat flux for both freestream gas mixtures. In both cases, the present simulations overpredict stagnation point convective heat flux compared with the experimental data. However, stagnation point convective heat flux values obtained in the present work are still inside the margin of error of the experimental data. The flow composed initially of \ch{CO2 + N2} yields an average stagnation point convective heat flux slightly lower than the flow composed purely of \ch{CO2}. Note that radiative phenomena are not included in the present analysis. Therefore, considering the analysis performed in this work, the impact of the Park's two-temperature model weight factors on the stagnation point convective heat flux is very small.

\section{CONCLUDING REMARKS}

The present research investigated reactive hypersonic flows under thermodynamic and chemical non-equilibrium conditions. The Navier-Stokes equations with source terms accounting for chemical reactions and non-equilibrium phenomena are solved. An 8-species chemical model simulates Mars' atmosphere. The solver employs a two-temperature model to represent the thermal non-equilibrium phenomena, using the translational-rotational and vibrational-electronic coupled temperature modes. The present work analyzed the influence of the weight factors of Park's two-temperature model on the flow behavior of Mars entry hypersonic flows. The results presented in this study are in terms of the Mach number, representing the shock wave position, temperature modes, representing the thermodynamic non-equilibrium phenomena, and the stagnation point convective heat flux. The results presented agree with experimental and numerical data available in the literature.

The present work aims to broaden the understanding of the influence of Park's two-temperature model weight factors on the flow behavior. 
The results presented in this analysis show that the weight factors impact the flow behavior. 
It is shown in this study that the shock position tends to move away from the vehicle body as the value of \( a \) increases. 
Moreover, the increase of the \( a \) weight factor value yields lower maximum \( T_{tr} \) and \( T_{ve} \) temperature mode values. 
The present work also shows that the stagnation point convective heat flux is essentially not affected by the variations of the weight factors. Variations in the stagnation point convective heat flux are of the order of \SI{0.15}{\percent} when considering all cases tested for a given atmosphere composition. 
Therefore, one can conclude that the changes caused by varying the weight factor values are negligible for the shock wave position and the stagnation point convective heat flux. 
However, the changes in the maximum temperature modes may not be negligible because the temperature dictates the chemical reaction rates, which, in turn, might affect the chemical composition and the radiative behavior of the mixture. 
These later effects, however, have not been addressed in the present study.

However, the present work cannot make any recommendations regarding the values of the weight factors other than those already proposed as good choices in the literature. Additional information on property distributions may be required to provide better calibration of the weight factor values. Nevertheless, this study demonstrated that the variations in the weight factor values yield changes in flow behavior that are consistent with the expected behavior of the theoretical formulation presented.

\section{ACKNOWLEDGEMENTS}

The authors gratefully acknowledge the support for the present research provided by Fundação de Amparo à Pesquisa do Estado de São Paulo, FAPESP, under the Scholarship Grant No.\ 2024/07590-2\@.
The authors acknowledge the availability of computational resources from the Center for Mathematical Sciences Applied to Industry, CeMEAI, funded by FAPESP under the Research Grant No.\ 2013/07375-0\@. 
The authors further acknowledge the National Laboratory for Scientific Computing, LNCC/MCTI, Brazil, for providing HPC resources of the SDumont supercomputer, under the HFWBTF project.
Support to the second author under the FAPESP Scholarship Grants No.\ 2021/02705-8 and 2022/07604-8, and additional support to the third author under FAPESP Research Grant No.\ 2013/07375-0 are also gratefully acknowledged.
Partial support for the present research was also provided by Conselho Nacional de Desenvolvimento Científico e Tecnológico, CNPq, under the Research Grant No.\ 315411/2023-6\@.
This study was financed in part by the Coordenação de Aperfeiçoamento de Pessoal de Nível Superior - Brasil (CAPES) - Finance Code 001\@.

\section{REFERENCES} 

\renewcommand{\refname}{}
\bibliography{bibfile.bib}

\end{document}